\documentclass[11pt,twoside]{article}
\usepackage[pdftex]{graphicx}
\usepackage{amsmath}
\usepackage{amssymb}
\usepackage{cite}

\usepackage{multicol}
\usepackage{epstopdf}

\begin{document}

\newcommand{\pst}{\hspace*{1.5em}}

\newcommand{\rigmark}{\em Journal of Russian Laser Research}
\newcommand{\lemark}{\em Volume 30, Number 5, 2009}

\newcommand{\be}{\begin{equation}}
\newcommand{\ee}{\end{equation}}
\newcommand{\bm}{\boldmath}
\newcommand{\ds}{\displaystyle}
\newcommand{\bea}{\begin{eqnarray}}
\newcommand{\eea}{\end{eqnarray}}
\newcommand{\ba}{\begin{array}}
\newcommand{\ea}{\end{array}}
\newcommand{\arcsinh}{\mathop{\rm arcsinh}\nolimits}
\newcommand{\arctanh}{\mathop{\rm arctanh}\nolimits}
\newcommand{\bc}{\begin{center}}
\newcommand{\ec}{\end{center}}

\renewcommand{\labelenumi}{(\alph{enumi})} 
\let\vaccent=\v 
\renewcommand{\v}[1]{\ensuremath{\mathbf{#1}}} 
\newcommand{\gv}[1]{\ensuremath{\mbox{\boldmath$ #1 $}}} 
\newcommand{\uv}[1]{\ensuremath{\mathbf{\hat{#1}}}} 
\newcommand{\abs}[1]{\left| #1 \right|} 
\newcommand{\avg}[1]{\left< #1 \right>} 
\let\underdot=\d 
\renewcommand{\d}[2]{\frac{d #1}{d #2}} 
\newcommand{\dd}[2]{\frac{d^2 #1}{d #2^2}} 
\newcommand{\pd}[2]{\frac{\partial #1}{\partial #2}} 
\newcommand{\pdd}[2]{\frac{\partial^2 #1}{\partial #2^2}} 
\newcommand{\pdc}[3]{\left( \frac{\partial #1}{\partial #2}
	\right)_{#3}} 
\newcommand{\ket}[1]{\left| #1 \right>} 
\newcommand{\bra}[1]{\left< #1 \right|} 
\newcommand{\braket}[2]{\left< #1 \vphantom{#2} \right|
	\left. #2 \vphantom{#1} \right>} 
\newcommand{\matrixel}[3]{\left< #1 \vphantom{#2#3} \right|
	#2 \left| #3 \vphantom{#1#2} \right>} 
\newcommand{\grad}[1]{\gv{\nabla} #1} 
\let\divsymb=\div 
\renewcommand{\div}[1]{\gv{\nabla} \cdot #1} 
\newcommand{\curl}[1]{\gv{\nabla} \times #1} 
\let\baraccent=\= 
\renewcommand{\=}[1]{\stackrel{#1}{=}} 

\thispagestyle{plain}

\label{sh}


\begin{center} {\Large \bf
		\begin{tabular}{c}
		THE PARTITION FORMALISM AND NEW\\[-1mm] ENTROPIC-INFORMATION INEQUALITIES\\[-1mm] FOR REAL NUMBERS ON AN EXAMPLE\\[-1mm] OF CLEBSCH-GORDAN COEFFICIENTS
		\end{tabular}
	} \end{center}

	\begin{center} {\bf
			Vladimir I.~Man'ko$^{1,2}$ and Zhanat Seilov$^{2,*}$
		}\end{center}
		
		\medskip
		
		\begin{center}
			{\it
				$^1$Lebedev Physical Institute, Russian Academy of Sciences\\
				Leninskii Prospect 53, Moscow, Russia 119991
				
				\smallskip
				
				$^2$Moscow Institute of Physics and Technology (State University)\\
				Institutskii per. 9, Dolgoprudnyi, Moscow Region Russia 141700
			}
			\smallskip
			
			$^*$Corresponding author e-mail:~~~seilov@live.ru\\
		\end{center}
\begin{abstract}\noindent
	We discuss the procedure of different partitions in the finite set
	of $N$ integer numbers and construct generic formulas for a
	bijective map of real numbers $s_y$, where $y=1,2,\ldots,N$,
	$N=\prod
	\limits_{k=1}^{n} X_k$, and $X_k$ are positive integers, onto the set of
	numbers $s(y(x_1,x_2,\ldots,x_n))$. We give the functions used to
	present the bijective map, namely, $y(x_1,x_2,...,x_n)$ and $x_k(y)$
	in an explicit form and call them the functions detecting the hidden
	correlations in the system. The idea to introduce and employ the
	notion of ``hidden gates'' for a single qudit is proposed. We obtain
	the entropic-information inequalities for an arbitrary finite set of
	real numbers and consider the inequalities for arbitrary
	Clebsch--Gordan coefficients as an example of the found relations
	for real numbers.
\end{abstract}

\medskip


\section{Introduction}  \pst
Recently some new entropic-information inequalities for a finite set
of nonnegative numbers~\cite{nonneg}, the matrix elements of unitary
matrices~\cite{marmo,mark1511} describing the probability
distributions, and particular examples for the Clebsch-Gordan
coefficients~\cite{art2} were obtained using specific partitions of
a set of integer numbers. In this article, for a set of $N$ real
numbers $s_y$ ($y=1,2,\ldots,N$) we introduce a set of nonnegative
numbers $p(y)=\dfrac{|s_y|}{\sum_{y'=1}^{N}|s_{y'}|}$. The set of
numbers $p(y)$ obtained can be interpreted as a probability
distribution with the properties presented in the form of entropic
equalities and inequalities.

The main goal of our work is to construct explicitly the functions
that provide the bijective map between the integer numbers
$y=1,2,\ldots,N$ ($N=\prod_{k=1}^n {X_k}$) and a set of variables
$x_k$ ($x_k=1,2,\ldots,X_k$). The function is used to describe
hidden correlations~\cite{physconf,hidden1,hiddenbell} in quantum
and classical systems that do not contain subsystems. The set of
values of these functions and their arguments have the geometrical
interpretation as the set of dots with integer nonnegative
coordinates $\{x_1,\ldots,x_n,y\}$ located on a hyperplane in the
$(n+1)$-dimensional space.

Using the function constructed, we can employ the
entropic-information inequalities known for joint probability
distributions to the case of arbitrary probability distributions of
one random variable and obtain new simple inequalities for an
arbitrary finite set of real numbers $s_y$. Specific examples of
such probability distributions are the modulus squared of matrix
elements of irreducible representations of classical Lie groups and
Clebsch--Gordan coefficients for the group representations.

This approach brings us to the other goal of the work, which is to
obtain new inequalities for special functions that determine the
matrix elements of the $SU(2)$-group irreducible representation and
the Clebsch--Gordan coefficients for quantum angular momentum.

This paper is organize as follows.

In Sec.~2, we present our motivation for constructing the function
providing the reversible map of integer numbers onto sets of
combinations (pairs, triples, etc.) of integers. In Sec.~3, we
obtain new inequalities for a finite set of real numbers and provide
new subadditivity and strong subadditivity
conditions~\cite{holevo,shiryaev,feller,kolmog}, known for composite
systems, for the probability distribution of one random variable
describing an indivisible system. In Sec.~4, for arbitrary
Clebsch--Gordan coefficients of the $SU(2)$-group we derive new
inequalities using the functions constructed. Finally, we give the
prospects and conclusions in Sec.~5.

\section{Partition of a Finite Set of Real Numbers} \pst
We denote by $s_y$ a set of $N$ real numbers that can be indexed by
the variable $y$, which takes the values $y=1, 2,\ldots,N$.

In this section, to represent the set of integer variables $y$ as
another set of $n$ integer variables $x_1,\ldots,x_n$, where
$x_i\in\{1,\ldots,X_i\}$, $i=1,\ldots,n$, i.e.,
$y=y(x_1,\ldots,x_n)$, we consider the partition of an integer
number $N$ introducing $n$ integer numbers $X_1,\ldots, X_n$.  For
distinctness, we assume $X_k\geq X_{k-1}$ for all $k\in(1,n)$. To
show such partitions, we construct bijective maps between integer
numbers $y$ and a set of integers $x_i$ ($i=1,\ldots,n$). These maps
are described by the function $y=y(x_1,\ldots,x_n)$ and $n$ reverse
functions $x_i=x_i(y)$, $i=1,\ldots,n$.

As an example, we construct the following bijective maps for the
partition of the integer number $N=8$. We consider this number as a
product of two factors ($N=4\times 2$). In this case, we introduce
the function of two variables $y=y(x_1,x_2)$. In the second case, we
consider this number as the product of three factors ($N=2
\times 2 \times 2$) and introduce the function of three variables $y=y(x_1,x_2,x_3)$.

The set of the values of the function $y$ is
\begin{center} $\begin{matrix} y & 1 & 2 & 3 & 4 & 5 & 6 & 7 & 8. \end{matrix}$ \end{center}
In the first case, the set of variables $(x_1, x_2)$ and $X_1=4$,
$X_2=2$ reads
\begin{center} $\begin{matrix} (x_1,x_2) & (1,1) & (2,1)& (3,1) & (4,1) & (1,2) & (2,2) & (3,2) & (4,2).\end{matrix}$\end{center}
In the second case, the set of variables $(x_1,x_2,x_3)$ reads
\begin{center} $\begin{matrix} (x_1,x_2,x_3) & (1,1,1) & (2,1,1) & (1,2,1) & (2,2,1) & (1,1,2) & (2,1,2) & (1,2,2) & (2,2,2).\end{matrix}$
\end{center}

Below we present the expressions describing bijective maps for
simple partitions of number $N=8$ into $2$ and $3$ factors along
with a generic case of the partition of integer number $N$ into $n$
factors.

\begin{itemize}
	\item
	The bijective map for the set of integer numbers $y$ and the set of two integer variables $x_1$ and
	$x_2$ reads
	\be \label{map21}    y(x_1,x_2)=x_1+(x_2-1)X_1,\qquad 1\leq x_1 \leq X_1,\quad 1\leq x_2 \leq X_2.\ee
	The function $x_1(y)$ is determined by the relation
	\be \label{map22}  x_1(y)=y \mod{X_1},\qquad 1\leq y \leq N.\ee
	The function $x_2(y)$ is determined by the relation
	\be \label{map23}  x_2(y)-1=\frac{y-x_1(y)}{X_1} \mod{X_2},\qquad 1\leq y \leq
	N.\ee In fact, we introduce the functions that provide the
	possibility to represent any probability distribution $f(y)$,
	$y=1,\ldots,N$ as a joint probability distribution $f(x_1,x_2)$ of
	the bipartite system.
	
	The relation of a pair of integers $(x_1,x_2)$ to an integer $y$ can
	also be illustrated, in view of the representation for the
	probability distribution of a composite system $AB$ by matrices
	$p(x_1,x_2)$ and $f(y)$ of the form
	
	{\small
		$\begin{pmatrix}
		p(1,1) & p(2,1) & \cdots & p(X_1,1)\\
		p(1,2) & p(2,2) & \cdots & p(X_1,2)\\
		\vdots & \vdots & \ddots & \vdots\\
		p(1,X_2) & p(2,X_2) & \cdots & p(X_1,X_2)
		\end{pmatrix} \equiv \begin{pmatrix}
		f(1) & f(2) & \cdots & f(X_1)\\
		f(1+X_1) & f(2+X_1) & \cdots & f(2 X_1)\\
		\vdots & \vdots & \ddots & \vdots\\
		f(1+X_1(X_2-1)) & f(2+X_1(X_2-1))x & \cdots & f(N)
		\end{pmatrix}$}.
	
	Here, the corresponding elements of these two matrices are
	numerically identical, e.g., $f(1)=p(1,1)$ and $f(N)=p(X_1,X_2)$.
	
	The joint-probability-distribution entropic-information inequalities
	show correlations (or ``hidden'' correlations~\cite{hiddenbell}) in
	bipartite systems, which is the reason to call the set of
	functions 
	$y(x_1,x_2)$, $x_1(y)$, and $x_2(y)$~(1)--(3)  the functions
	detecting hidden correlations.
	
	Equation (\ref{map21}) can be interpreted as the equation of a plane
	in three-dimensional space with coordinates $(x_1, x_2,y)$ in the
	domain restricted by values of $X_1$ and $X_2$. This plane can also
	be determined by the equation $(\vec{n},\vec{r}-\vec{r_0})=0$, where
	the vector $\vec{n}$ is a normal to the plane, and the vector
	$\vec{r}-\vec{r_0}$ is an arbitrary vector on the plane.
	
	In Fig.~1, we present the example of the plane $y(x_1,x_2)$ with
	$X_1=X_2=4$ defined by the equation
	$$
	y=y(x_1,x_2)=x_1+(x_2-1)4, \qquad 1\leq x_1,x_2 \leq 4,
	$$
	or by the normal vector $\vec{n}=\{1,4,-1\}$ and vector
	$\vec{r_0}=\{1,1,1\}$. The domain $\{1\leq x_1\leq 4,~1\leq x_2
	\leq 4 \}$ represents a parallelogram in three-dimensional space
	with coordinates $(x_1,x_2,y)$ with vertices in dots
	$\{1,1,1\},~\{1,4,5\},~
	\{4,1,12\}$, and $\{4,4,16\}$.
	
	In Figs.~2 and 3, we present the plots of functions $x_1(y)$ and
	$x_2(y)$ describing the bijective map~(\ref{map22}) and
	(\ref{map23}).
	
	\begin{figure}[ht]
		\begin{minipage}[t]{67mm}
			\includegraphics[width=67mm]{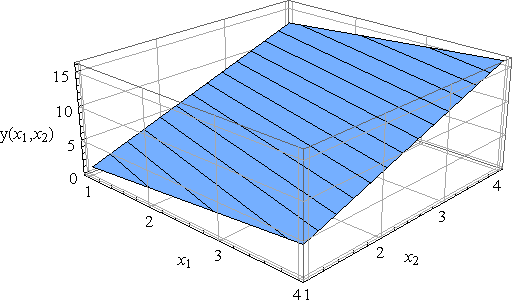}
			\caption{The plane $(\vec{n},\vec{r}-\vec{r_0})=0$ with domain
				$1\leq x_1,~x_2 \leq 4$, normal vector $\vec{n}=\{1,4,-1\}$, and
				vector $\vec{r_0}=\{1,1,1\}$.}
		\end{minipage}
		\hspace{1mm}
		\begin{minipage}[t]{51mm}
			\includegraphics[width=51mm]{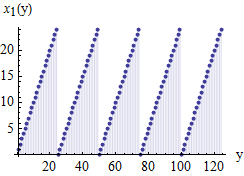}
			\caption{
				The coordinate $x_1$ versus the integer $y$; here $N=125$, $X_1=25$,
				and $X_2=5$. }
		\end{minipage}
		\hspace{1mm}
		\begin{minipage}[t]{51mm}
			\includegraphics[width=51mm]{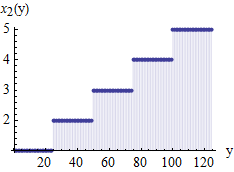}
			\caption{
				The coordinate $x_2$ versus the integer $y$; here $N=125$, $X_1=25$,
				and $X_2=5$. }
		\end{minipage}
	\end{figure}
	
	\item
	The bijective map for the set of integer numbers $y$ and the set of
	three integer variables $x_1$, $x_2$, and $x_3$ is described by
	functions $y(x_1,x_2,x_3)$, $~x_1(y)$, $~x_2(y)$, and $x_3(y)$,
	where
	\be \label{map31}
	y=y(x_1,x_2,x_3)=x_1+(x_2-1)X_1+(x_3-1)X_1 X_2,\quad 1\leq x_i \leq X_i,
	~i\in[1,3].\ee
	The function $x_1(y)$ is determined by the relation
	\be \label{map32} x_1(y)=y\mod{X_1}. \ee
	The function $x_2(y)$ reads
	\be \label{map33} x_2(y)-1=\frac{y-x_1(y)}{X_1}\mod{X_2}. \ee
	The function $x_3(y)$ is
	\be \label{map34} x_3(y)-1=\frac{y-x_1(y)-(x_2(y)-1)X_1}{X_1 X_2}\mod{X_3}. \ee
	The domain for functions $x_1(y)$, $x_2(y)$, and $x_3(y)$ is given
	as $1\leq y ~\leq N$ in all three cases.
	
	\item
	The bijective map for the set of integer numbers $y$ and the set of
	$n$ integer variables $x_i$ reads
	\be
	y=y(x_1,x_2,\ldots,x_n)=x_1+\sum_{k=2}^{n}(x_k-1)\prod_{j=1}^{k-1}X_j, \qquad
	1\leq x_i \leq X_i,\quad i\in[1,n].        \label{su}\ee
	The function $x_k(y)$ is
	\be x_k(y)-1=\frac{y-(x_1+\sum_{i=2}^{k}(x_i-1)\prod_{j=1}^{i-1}X_j)}{\prod_{j=1}^{k}X_j}\mod{X_k},
	\qquad k=1,\ldots,n, \quad 1\leq y \leq N. \label{us}
	\ee
	For example, $x_4(y)=\dfrac{y-(x_1+(x_2-1)X_1+(x_3-1)X_1 X_2)}{X_1
		X_2 X_3}\mod{X_4}$.
\end{itemize}
\noindent
Equation~(\ref{su}) determines an $n$-dimensional plane in an
($n+1$)-dimensional space with coordinates $(x_1,x_2,\ldots,y)$. The
other way to determine this plane is to use the equation
$$(\vec{n},\vec{r}-\vec{r_0})=0,\qquad
\vec{n}=\{1,X_1,X_1X_2,\ldots,X_1\ldots\ldots\ldots X_{n-1},-1\};$$
$\vec{r_0}$ is the vector determining the position to any dot on the
plane; for example, it may be the vector
$\vec{r_0}=\{1,1,\ldots,1\}$.

Using the above formulas~(\ref{su}) and (\ref{us}), we can find $y$
if we know $x_1, x_2,\ldots,x_n$; vice versa, if we know $y$, we can
find every $x_i$, $i=1,\ldots,N$. In other words, we have the
one-to-one correspondence between the set of integers $y$ and the
set of integers $x_i$, and the map can be illustrated on an example
of the above partition into two factors of number $N=X_1\times X_2 =
4\times4$, where the integer number $y$ corresponds to the set of
two integer numbers $x_1$ and $x_2$.

In Fig.~4, we illustrate the intersection between the plane
$y(x_1,x_2)$ determined in Eqs.~(1)--(3)    
and the plane $y=5$ or the intersection between the plane with
normal vectors $\vec{n_1}=\{1,4,-1\}$, $\vec{n_2}=\{0,0,1\}$ and the
common-position vector $\vec{r_0}=\{2,2,5\}$. This intersection is a
line on which only one dot with both integer numbers $x_1$ and $x_2$
is located; in this case, according to (1)--(3), 
$x_1=2$ and $x_2=2$. This intersection is a line determined by the
system of equations for the plane $y=5$ and the plane shown in
Fig.~1 and determined by Eqs.~(1)--(3). 
These equations for the particular example read
\be
\begin{cases}
	& y-5=0,\\
	& y-x_1- 4(x_2-1)=0.
\end{cases}
\ee
Also the intersection line is determined by the equation
$[\vec{a},\vec{r}-\vec{r_0}]=\vec{0}$, where the vector
$\vec{a}=[\vec{n_1},\vec{n_2}]=\{4,1,0\}$ and the vector
$\vec{r_0}=\{2,2,5\}$ provide the position of the dot on the line.

For every integer number $y'$, we plot the intersection (line) of
the plane $y=y'$ with the plane $y(x_1, x_2)$ determined by
(1)--(3) 
and then project the lines to the plane parallel to the plane $y=0$.
Every such intersection corresponds to only one dot with integers
$x_1$ and $x_2$ within the domain $x_1 \in [1,X_1]$ and $x_2\in
[1,X_2]$.

In the case presented, we have 16 intersections between the
mentioned plane $y=y(x_1,x_2)$ and planes $y=y'$ for every integer
$y'=1,\ldots,16$. Every intersection corresponds to one set of three
integer numbers $y$, $x_1$, and $x_2$. These intersections, being
projected into one plane, have the same angle $\alpha$, with
$\tan\alpha=\dfrac{X_2-1}{X_2}\,\dfrac{1}{X_1}$. We show the
projections in Fig.~5.

In $(n+1)$-dimensional space, the intersection of two planes with
normal vectors $\vec{n_1}$ and $\vec{n_2}$ is a ($n-1$)-dimensional
plane determined by the equation
$[\vec{a},\vec{r}-\vec{r_0}]=\vec{0}$ with the vector
$\vec{a}=[\vec{n_1},\vec{n_2}]$ and the position vector $\vec{r_0}$.

\begin{figure}[ht]
	\begin{minipage}[t]{86mm}
		\centerline{\includegraphics[width=70mm]{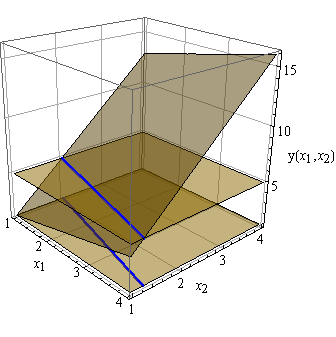}}
		\caption{Intersection between the plane\newline $y(x_1,x_2)=x_1+4(x_2-1),~
			1\leq x_1,x_2\leq 4$\newline and the plane $y=5$.}
		\label{figLeft}
	\end{minipage}
	\hspace{1mm}
	\begin{minipage}[t]{86mm}
		\centerline{\includegraphics[width=63mm]{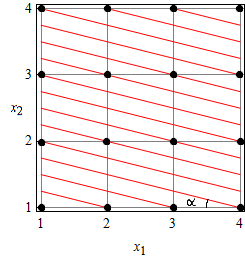}}
		\caption{Projections of intersections between the plane
			$y(x_1,x_2)=x_1+4(x_2-1),~1\leq x_1,x_2\leq 4$ and planes $y=y'$,
			where $y'=1,\ldots,16$.}
		\label{figRight}
	\end{minipage}
\end{figure}

\section{Properties of the Shannon Entropy in Terms of the Probability \textit{p(y)}}

\subsection{Subadditivity Condition}    \pst
One of the main properties of the Shannon entropy for the joint
probability distribution of a bipartite system $AB$ is the
subadditivity condition, which reads
\be \label{sub}
H(A)+H(B) \geq H(AB),
\ee
where $A$ and $B$ are the subsystems of the system $AB$. Below we
consider coordinates $x_1$ and $x_2$ as discrete integer variables.

Let $p(x_1,x_2)$, where $1\leq x_1 \leq X_1$, $~1\leq x_2 \leq X_2$,
be a joint probability distribution for a composite system $AB$ with
finite number $N=X_1 X_2$ of elements,
$\mathcal{P}(x_1)=\sum_{x_2=1}^{X_2}p(x_1,x_2)$ be the probability
distribution for system $B$, and $\Pi=\sum_{x_1=1}^{X_1}p(x_1,x_2)$
be the probability distribution for system $A$. Now we can rewrite
the subadditivity condition~(\ref{sub}) as follows:
\be     \label{sub2}
-\sum_{x_1=1}^{X_1}\mathcal{P}(x_1)\log\mathcal{P}(x_1)
-\sum_{x_2=1}^{X_2}\Pi(x_2)\log\Pi(x_2) \geq
-\sum_{x_1=1}^{X_1}\sum_{x_2=1}^{X_2} p(x_1,x_2)\log p(x_1,x_2).
\ee
The transition from the probability distribution $p(x_1,x_2)$,
$1\leq x_1 \leq X_1$, $~1\leq x_2 \leq X_2$ depending on two integer
variables to the probability distribution $f(y)$, $1\leq y
\leq N$ depending on one integer variable $y$ can be performed
using the bijective map~(\ref{map21}); we have the probability
distribution of the indivisible system
\be f\big (y(x_1,x_2)\big)=p(x_1,x_2).\ee
The two marginal probability distributions read
\be \mathcal{P}(x_1)=\sum_{x_2=1}^{X_2}p(x_1,x_2)=\sum_{x_2=1}^{X_2}f(y(x_1,x_2))
=\sum_{x_2=1}^{X_2}f(x_1+(x_2-1)X_1) \ee
and
\be \Pi(x_2)=\sum_{x_1=1}^{X_1}p(x_1,x_2)=\sum_{x_1=1}^{X_1}f(y(x_1,x_2))=
\sum_{x_1=1}^{X_1}f(x_1+(x_2-1)X_1). \ee
This transition provides the possibility to rewrite inequality
(\ref{sub2}) in the form
\bea
-\sum_{x_1=1}^{X_1} \left(\sum_{x_2=1}^{X_2} f(x_1+(x_2-1)X_1)
\log\sum_{x_2=1}^{X_2}
f(x_1+(x_2-1)X_1)\right)\nonumber\\
-\sum_{x_2=1}^{X_2} \left(\sum_{x_1=1}^{X_1} f(x_1+(x_2-1)X_1)
\log\sum_{x_1=1}^{X_1} f(x_1+(x_2-1)X_1)\right) \geq
-\sum_{y=1}^{N} f(y)\log f(y). \label{explsub}
\eea
Inequality (\ref{explsub}) is the main result of our consideration
of a set of nonnegative numbers
$p(y)=\dfrac{|s_y|}{\sum_{y'=1}^{N}|s_{y'}|}$ associated with real
numbers $s_y$. Such set $p(y)$, considered as the probability
distribution and denoted in this case as $f(y)$, always satisfies
the obtained inequality (\ref{explsub}).

The difference between the left-hand side and the right-hand side of
inequality~(\ref{explsub}) gives the Shannon information
$I=H(A)+H(B)-H(AB)\geq 0$, or joint information, which takes only
nonnegative values.

\subsection{Entropy of the Composite System}    \pst
The Shannon entropy of a composite system with subsystems $A_1
A_2\ldots A_n$ satisfies the equality
\be \label{conditional}
H(A_1A_2\ldots A_n)= H(A_1)+H(A_2|A_1)+\cdots+
H(A_n|A_1,A_2,\ldots,A_{n-1}).
\ee
Here, the Shannon entropy of the conditional probability
distribution $Q(x_1|x_2)=\dfrac{p(x_1,x_2)}{\Pi(x_2)}$ is
$$
H(A|B)=H(Q(x_1|x_2))=-\sum_{x_1=1}^{X_2}\sum_{x_2=1}^{X_1}p(x_1,x_2)\log
Q(x_1,x_2).
$$
In the case of composite system $AB$, expression~(\ref{conditional})
reduces to the form
\be
H(AB)=H(B)+H(A|B)\qquad \mbox{or}\qquad
H(p(x_1,x_2))=H(\Pi(x_2))+H(Q(x_1|x_2)).\label{cond}
\ee
Using the bijective map described by Eqs.~(1)--(3), 
we present Eq.~(\ref{cond}) in terms of the function $f(y)$ as
follows:
\begin{eqnarray}
	-\sum_{y=1}^{X_1 \cdot X_2} f(y)\log а(y)&=&
	-\sum_{x_2=1}^{X_2} \sum_{x_1=1}^{X_1} f(x_1+(x_2-1)X_1)\log \sum_{x_1=1}^{X_1} f(x_1+(x_2-1)X_1)
	\nonumber\\
	&&-\sum_{x_1=1}^{X_2}\sum_{x_2=1}^{X_1} f(x_1+(x_2-1)X_1) \log \frac{f(x_1+(x_2-1)X_1)}{\sum_{x_1=1}^{X_1}f(x_1+(x_2-1)X_1)},
\end{eqnarray}
where $f(y)$ is the probability distribution of one random variable
$1\leq y \leq N$. Here, the integers $x_1$ and $x_2$ belong to
domain $1\leq x_1 \leq X_1$ and $1\leq x_2 \leq X_2$.

\section{Entropic Inequalities for Clebsch--Gordan Coefficients} \pst
The Clebsch--Gordan coefficients $\braket{j_1 m_1 j_2 m_2}{jm}$ are
defined as follows~\cite{LL}:
\be
\psi_{jm}=\sum_{m_1,m_2}\braket{j_1 m_1 j_2 m_2}{jm} \psi_{j_1 m_1}^{(1)}
\psi_{j_2 m_2}^{(2)},  \qquad m=m_1+m_2,
\ee
where $\psi_{jm}$ is the wave function of the spin system with spin
$j$ and the spin projection $m$, and $\psi_{j_1 m_1}^{(1)}$ and
$\psi_{j_2 m_2}^{(2)}$ are two wave functions of the spin system
with spin $j_1$ and the spin projection $m_1$ and spin $j_2$ and the
spin projection $m_2$, respectively.

\subsection{Subadditivity Inequality}    \pst
We can express the Clebsch--Gordan coefficients as functions of
integer variable $y=1,\ldots,N$, where $N=(2j_1+1)(2j_2+1)$, using
the relation $m_i=x_i-j_i-1$,  $i=1,2$ and the relation between
$x_i$ and $y$ given by (1)--(3). 
We employ the partition $N=X_1\times X_2$, where $X_i=2j_i+1$
($i=1,2$).

We adopt the approach~\cite{art2} and apply it to the squares of the
Clebsch--Gordan coefficients $\braket{m_1,m_2}{jm}$ as the
probabilities
\be
p(m_1,m_2)=|\braket{m_1,m_2}{jm}|^2=f(y(m_1,m_2)).
\ee
In the formulas below, we omit arguments $j$ and $m$ in the
definition of probabilities connected with the Clebsch--Gordan
coefficients, since for such the probabilities the values of $j$ and
$m$ are fixed.

In view of (1)--(3), 
we obtain the functions $m_1(y)$ and $m_2(y)$ as follows:
$$
m_1(y)=[y\mod(2j_1+1)]-j_1-1,\qquad
m_2(y)=\left[\frac{y-[y\mod(2j_1+1)]}{2j_1+1}\mod (2j_2+1)\right]-j_2.
$$
We introduce the probability mentioned above in terms of the integer
number $y$; it reads
\be \label{ps}
f(y)=\left|\braket{[y\mod(2j_1+1)]-1-j_1,\left[ \frac{y-[y\mod(2j_1+1)]}{2j_1+1}\mod{(2j_2+1)}\right]-j_2}{jm}\right|^2.
\ee
In \cite{art2}, we derived the subadditivity condition~(\ref{sub})
for the Shannon entropy in terms of the Clebsch--Gordan
coefficients, which reads
\bea
-\sum_{m_2=-j_2}^{j_2}|\braket{j_1 m_1 j_2 m_2}{jm}|^2
\log\left[\sum_{m_2=-j_2}^{j_2}|\braket{j_1 m_1 j_2 m_2}{jm}|^2 \right]\nonumber  \\
-\sum_{m_1=-j_1}^{j_1}|\braket{j_1 m_1 j_2 m_2}{jm}|^2
\log\left[\sum_{m_1=-j_1}^{j_1}|\braket{j_1 m_1 j_2 m_2}{jm}|^2\right]\nonumber\\
\geq -\sum_{m_1=-j_1}^{j_1} \sum_{m_2=-j_2}^{j_2}|\braket{j_1 m_1 j_2 m_2}{jm}|^2
\log\left[\sum_{m_1=-j_1}^{j_1} \sum_{m_2=-j_2}^{j_2}|
\braket{j_1 m_1 j_2 m_2}{jm}|^2\right].  \label{ccgsa}
\eea
This inequality is a new entropic inequality for the Clebsch--Gordan
coefficients.

After combining expressions (\ref{ps}) and (\ref{ccgsa}), we arrive
at the subadditivity condition~\cite{lieb, petz} in terms of a
single variable $ y $,
\bea
&&-\sum_{m_2=-j_2}^{j_2}f(y(m_1,m_2))\log\sum_{m_2=-j_2}^{j_2}f(y(m_1,m_2))
-\sum_{m_1=-j_1}^{j_1}f(y(m_1,m_2)) \log\sum_{m_1=-j_1}^{j_1}f(y(m_1,m_2)) \nonumber \\
&&\geq -\sum_{y=0}^{N}f(y(m_1,m_2)) \log
\sum_{y=0}^{N}f(y(m_1,m_2)).
\eea

\subsection{Strong Subadditivity Condition} \pst
The strong subadditivity property for Shannon entropy states that
the conditional information for tripartite system $ABC$ takes
nonnegative values and reads
\be \label{ssa}
H(ABC)+H(B) \geq H(AB) + H(BC).
\ee
In view of the one-to-one correspondence between the integer
variable $y$ and a pair of integer variables $(x_1, x_2)$ and
between $y$ and a triple of integer variables $(t_1,t_2,t_3)$, i.e.,
using formulas~(1)--(3) 
and (4)--(7),  
we can construct a bijective map between $(x_1, x_2)$ and
$(t_1,t_2,t_3)$. This means that we can write the strong
subadditivity condition for the quantum system of two spins; it is
\begin{eqnarray*}
	x_1(t_1,t_2,t_3)=y\mod{X_1}=\big[(t_3-1)T_2 T_1 + (t_2-1)T_1+t_1\big]\mod{X_1},\\
	x_2(t_1,t_2,t_3)-1=\frac{y-x_1}{X_1}\mod{X_2}=\frac{\big[(t_3-1)T_2 T_1 +
		(t_2-1)T_1+t_1\big]-x_1}{X_1}\mod{X_2},
\end{eqnarray*}
with the domain $1\leq t_i \leq T_i$, $~i=1,2,3$ for functions
$x_1(t_1,t_2,t_3)$ and $x_2(t_1,t_2,t_3)$.

Now we are in the position to rewrite the squares of the
Clebsch--Gordan coefficients as a probability distribution $g$
depending on three integer variables $(t_1,t_2,t_3)$, which take
values $N=T_1 T_2 T_3$, $~1 \leq t_i\leq T_i$, $~i=1,2,3$ instead of
two indices $m_1$ and $m_2$; it is
\begin{eqnarray} \label{p3}
	\resizebox{0.96\textwidth}{!}{$
		p\big(m_1(t_1,t_2,t_3),m_2(t_1,t_2,t_3)\big)=g(t_1,t_2,t_3)
		=\Big|\Big\langle t_1+(t_2-1)T_1+(t_3-1)T_1 T_2\,\mod(2j_1+1)-1-j_1,$} \nonumber\\
	\resizebox{0.95\textwidth}{!}{$\frac{[t_1+(t_2-1)T_1+(t_3-1)T_1 T_2]
			-\big[t_1+(t_2-1)T_1+(t_3-1)T_1 T_2\big(\mod(2j_1+1)\big)\big]}{2j_1+1}
		\mod(2j_2+1)-j_2\mid{jm}\Big\rangle\Big |^2.$}\nonumber\\
\end{eqnarray}
Combining the strong subadditivity (\ref{ssa}) and the expression
for Clebsch--Gordan coefficients as probability distributions $f(y)$
and $g(t_1,t_2,t_3)$~(\ref{p3}), we obtain the following inequality:
\begin{multline}
	-\sum_{y=1}^{N}f(y)\log f(y) -\sum_{t_2=1}^{T_2}
	\left(\sum_{t_1=1}^{T_1} \sum_{t_3=1}^{T_3} g(t_1,t_2,t_3)
	\log\sum_{t_1=1}^{T_1} \sum_{t_3=1}^{T_3}g(t_1,t_2,t_3) \right) \\
	\geq -\sum_{t_1=1}^{T_1}\sum_{t_2=1}^{T_2}
	\left(\sum_{t_3=1}^{T_3} g(t_1,t_2,t_3) \log\sum_{t_3=1}^{T_3} g(t_1,t_2,t_3) \right)
	-\sum_{t_2=1}^{T_2}\sum_{t_3=1}^{T_3}
	\left(\sum_{t_1=1}^{T_1} g(t_1,t_2,t_3) \log\sum_{t_1=1}^{T_1} g(t_1,t_2,t_3) \right).
\end{multline}
This inequality is a new relation for the Clebsch--Gordan
coefficients.

\section{Conclusions}   \pst
We obtained functions (\ref{su}) and (\ref{us}) for the bijective
map between the sequence of integer numbers and the sequence of sets
of $n$ integer numbers and called them the function detecting the
hidden correlations. These functions give a simple technique to
consider any indivisible system consisting of a finite number of $N$
elements with probabilities as a system consisting of $n$
subsystems; this fact allows us to apply the properties of
multipartite systems to indivisible systems. We employed this
technique to obtain the properties of the Shannon entropy for
probabilities associated with distributions, such as the
subadditivity and strong subadditivity conditions for a single
indivisible system.

The technique elaborated can be applied to any set of real numbers,
which we associate with a set of probabilities according to the
formula $p(y)=\dfrac{|s_y|}{\sum_{y'=1}^{N}|s_{y'}|}$, where $p(y)$
is the probability associated with a real number $s_y$, and $y$
takes integer values from $1$ to the number $N$ of elements in the
set of considered real numbers. We also applied the new technique to
obtain new entropic inequalities for the Clebsch--Gordan
coefficients, which were considered as probability distributions of
one random variable.

For an arbitrary $N$$\times$$N$ matrix $A$, there exists the matrix
$\rho=\dfrac{A^\dag A}{\mbox{Tr}\,{A^\dag A}}$ with the properties
of the density matrix, i.e., $\rho^\dag = \rho$, Tr$\,\rho=1$, and
$\rho\geq 0$. Using the entropic-information inequalities known for
density matrices, one can obtain the corresponding inequalities for
arbitrary $N$$\times$$N$ matrices $A$.

The results obtained in this work for real numbers and probability
distributions will be extended in a future publication to the
properties of arbitrary matrices. The entropic-information
inequalities obtained can be applied~\cite{glush, kikt1, kikt2} in
experiments with superconducting qudits~\cite{shalibo, pashkin,
	devoret} based on the Josephson junction discussed in \cite{SovLRes,
	zeillinger}. The obtained explicit formulas of the introduced
functions can be used to discuss the properties of quantum
correlations like the violation of Bell inequalities~\cite{bell},
contextuality problems~\cite{klyachko, rastegin}, entanglement
criteria~\cite{peres,horodecki}, and other correlations discussed
in~\cite{petz2,bengtsson} for different systems. Also it is worth
noting that functions~(1)--(3) in different notation were presented
in the PhD Thesis~\cite{pasquale} and employed in \cite{facchi}.

In the future publication, we relate the introduced formalism to the
star-product quantization scheme discussed, e.g., in
\cite{olga-star,PatriMPLA} and study a possibility to introduce and
employ the notion of ``hidden states'' for single qudit states (see
\cite{Kiktenko}) extending the notion of gates for composite systems
to the case of noncomposite systems.

\section*{Acknowledgments}  \pst
The formulation of the problem of ``hidden states'' and the results
of Sec.~3 are due to V. I. Man'ko, who is supported by the Russian
Science Foundation under Project No.~16-11-00084; the work was
performed at the Moscow Institute of Physics and Technology.

\end{document}